\documentclass[english,aps,preprint]{revtex4-1}
\usepackage[T1]{fontenc}
\usepackage[utf8]{luainputenc}
\usepackage{amsmath}
\usepackage{amssymb}
\usepackage{graphicx}
\usepackage{wasysym}
\usepackage{esint}

\makeatletter
\@ifundefined{showcaptionsetup}{}{%
 \PassOptionsToPackage{caption=false}{subfig}}
\usepackage{subfig}
\makeatother

\usepackage{babel}
\begin{document}

\title{Real space renormalization group for Ising Spin Glass and other glassy
models. \\
I. Disordered Ising model. General formalism}

\author{A.N. Samukhin}

\email{alex.samukhin@mail.ru}

\affiliation{A.F. Ioffe Physical \& Technical Institute, 194021, Polytechnicheskaya
26, St. Petersburg, RUSSIA}

\date{\today}
\begin{abstract}
Here is the first part of the summary of my work on random Ising model
using real-space renormalization group (RSRG), also known as a Migdal-Kadanoff
one. This approximate renormalization scheme was applied to the analysis
thermodynamic properties of the model, and of probabilistic properties
of a pair correlator, which is a fluctuating object in disordered
systems.
\end{abstract}



\maketitle

\section{Introduction\label{sec:Intro}}

The main goal of this series of publications is to introduce and apply
a new theoretical method aimed at the description of the wide class
of statistical phenomena known under the name ''spin glasses'' \citep{ea75}.
The name originates from the specific class of disordered magnetics
with elementary dipoles (spins) being frozen in random orientations.
It appeared, however, that similar phenomena are widespread in Nature.
Numerous phenomena in physics (amorphous magnetism and ferroelectricity,
structural glasses), biology (patterns formation in neural networks,
proteins folding), and computer science (storage and interaction of
patterns in memory) may be described using a mathematical models of
the same class. Numerous books and reviews on the subject are available
(see \citep{mpv87,by86,vsd93,fh93,cc05,fz10,alg13,hn01} and references
cited therein).

Among all the models, describing glassy behavior, the simplest is
the disordered Ising model with competing interactions. Originally
it was introduced to describe glassy state of a solution of a magnetic
metal in a nonmagnetic one, e.g. iron in gold. There the point is
the exchange interaction between magnetic ions may be either ferro-
or antiferromagnetic, depending upon the distance between ions, which
is a fluctuating random variable. The specific feature of such a system
is the possibility of \emph{frustrations} in it. A primitive illustration
may looks like this: Let us consider three Ising spins, $\sigma_{1,2,3}=\pm1$.
The Hamiltonian is:
\[
{\cal H}=-J_{12}\sigma_{1}\sigma_{2}-J_{23}\sigma_{2}\sigma_{3}-J_{31}\sigma_{3}\sigma_{1}\,.
\]
If all the pair interaction energies $J_{ij}>0$ (purely ferromagnetic
case), then the ground state is two-fold degenerate, either all spins
are $+1$, or $-1$. However, if one or all three interactions are
negative, then the specific case becomes possible. For example, let
us set $J_{12}=J_{23}=J>0,\,J_{31}=-J$. Spin configurations $\left(+++\right)$,
$\left(-++\right)$ and $\left(++-\right)$ have the energy, equal
to $-J$, while the configuration $\left(+-+\right)$ has the energy
$+3J$. The ground state becomes six-fold degenerate. The situation
is quite similar if $J_{12}=J_{23}=J_{31}=-J<0$. This reasoning can
be easily generalized to the case, when $N$ Ising spins are arranged
in a ring, and the number of antiferromagnetic interactions is odd.
The special attention should be paid to the fact, that, strictly speaking,
this degeneracy exists only if all the interaction energies are equal
in value, $\left|J_{ij}\right|=J$. For example, if in the triangle
we have: $J_{12}=J_{23}=J$, $J_{31}=-\tilde{J}$, and $J>\tilde{J}>0$,
the ground state is ferromagnetic and only two-fold degenerate. If
to assume $J_{ij}$ to be fluctuating random variables, then, in a
common case, the measure of frustrated configurations is zero. 

If there is a macroscopic number of frustrations in a macroscopic
system of a size $N\to\infty$, then the degeneracy of the ground
state appears to be $\sim a^{N}$, $a\gtrsim1$. This leads to a nonzero
value of the entropy in the thermodynamic limit \citep{fss13}. The
example is the antiferromagnetic Ising model on the triangular lattice.
However, in a disordered systems frustrations, strictly speaking,
exist only if the distribution of couplings is something very special,
e.g. $p\left(J_{ij}\right)=c\delta\left(J_{ij}-J\right)+\left(1-c\right)\delta\left(J_{ij}+J\right)$.
Usually, some configurations \emph{close} to frustration may be realized,
rather then true frustration, if ferro- and antiferromagnetic interactions
coexist in a system. Therefore, one can expect the residual entropy
to be zero, but some kind of singularity of the entropy and of the
specific heat at zero temperature is possible.

Originally the Migdal-Kadanoff real space renormalization group (RSRG)
was introduced more then fifty years ago by Leo Kadanoff \citep{lpk66}.
Then its continuous rescaling version was introduced by A.A. Migdal
\citep{aam75}. Comprehensive review of further developments of the
RSRG approach was recently given in \citep{ewkk14}.

The first attempt to apply RSRG to spin glass problem was made by
Droz and Matasinas \citep{dm80}. However, there severe and unjustified
approximations were made for the problem to be solvable. Then the
RSRG with integer rescaling factor was used in a number of works (see
e.g. \citep{th96,ncnc99,mbd98,mprz99,sn09}) The RSRG analysis was
applied in \citep{onm08} to analyze the multicritical points (coexistence
of (para/ferro)magnetic and spin glass phases) positions. A different
version of RSRG for spin glasses was used in \citep{cp11,mc11}. Tensor
RSRG for spin glasses was used in \citep{wqz14}, and RSRG for Ising
model with long range coupling was elaborated in \citep{cm16}

A functional version of RSRG with continuous rescaling was introduced
in \citep{spjw98} to study the behavior of conductivity at and near
the percolation threshold. Actually the RSRG is exact for one-dimensional
systems with short-range interaction, so it was applied for $\left((1+\epsilon\right)$-dimensional
fractals.It was a surprise, that renormalization equations (nonlinear
integro-differential ones) there allow exact solution, provided by
the saddle point approximation in the integrals.

This paper, first in a series, is devoted to the construction of mathematical
formalism of RSRG for a disordered Ising model. In Section \ref{sec:Statistical}
general concepts of the statistical mechanics of disordered systems
was applied to the Ising model. The crucial point is to perform \emph{configuration
average }of the free energy. Here a new representation of the distribution
of interaction energies is introduced and named ${\cal T}$-representation.
Formulas, connecting ${\cal T}$-representation with Fourier (or two-sided
Laplace) one are derived. Averaged free energy is defined as a \emph{functional
of the distribution of interaction energies.} Adding one extra link,
not necessary between nearest neighbors, and calculating the corresponding
variation of the free energy as a function of the strength the length
of this link provide us with \emph{generating function for the momenta
of the spin-spin correlator}, --- this is a straightforward generalization
of the approach elaborated by A.N. Vasil'ev \citep{anv98}.

In Section \ref{sec:RSRG} RSRG for a disordered Ising model is formulated,
first in Subsection \ref{sub:hierarchical} --- for a hierarchical
graph. Here the rescaling factor is integer. In Subsection \ref{sub:Infinitezimal}
the transition to infinitesimal rescaling is considered. In Subsection
\ref{sub:Free} explicit expressions for thermodynamic quantities
are given. The Section \ref{sec:Conclusions} contains brief summary
of results. Finally, in the last Section \ref{sec:Announcement} some
preliminary results of the next paper are adduced. Technical details
are taken into two Appendices.

\section{Statistical mechanics of the disordered Ising model. \label{sec:Statistical}}

\subsection{The Ising model.\label{sub:Ising}}

\subsubsection{Definitions.\label{sub:definitions}}

The Ising model consists of $N$ binary variables (``spins'') $\sigma_{i}$,
$i=1,2,\dots N$, that can be in one of two states, $\sigma_{i}=\pm1$,
or ``spin up/down''. These spins are either placed randomly in a
D-dimensional space, or arranged in a graph, usually a D-dimensional
lattice. The Hamiltonian of the model is:
\begin{equation}
\mathcal{H}\left(\left\{ \sigma_{i}\right\} \right)=-\sum_{i=1}^{N}h_{i}\sigma_{i}-\frac{1}{2}\sum_{i,j=1}^{N}J_{ij}\sigma_{i}\sigma_{j}.\label{eq:100}
\end{equation}
Here $h_{i}$ are local magnetic fields. Here I will assume $h_{i}=0$.
$J_{ij}=J\left(\mathbf{r}_{i}-\mathbf{r}_{j}\right)$ are pair interaction
energies. They may be either ferromagnetic (FM), $J_{ij}>0$, or antiferromagnetic
(AFM), $J_{ij}<0$, which favors two spins in the pair to be either
aligned, $\sigma_{i}=\sigma_{j}$, or anti-aligned, $\sigma_{i}=-\sigma_{j}$.

The partition function of the model is:
\begin{equation}
Z_{N}\left(\left\{ K_{ij}\right\} \right)=\sum_{\left\{ \sigma_{i}=\pm1\right\} }\exp\left(\frac{1}{2}\sum_{i,j}^{N}K_{ij}\sigma_{i}\sigma_{j}\right)\,.\label{eq:110}
\end{equation}
Here $K_{ij}=J_{ij}/T$ are the reduced coupling energies ($T$ is
the temperature in the energy units). The free energy $\mathcal{F}$
of the system is:
\begin{equation}
\mathcal{F}_{N}\left(\left\{ K_{ij}\right\} \right)=-T\ln Z_{N}\left(\left\{ K_{ij}\right\} \right)\,.\label{eq:120}
\end{equation}
The pair correlator is defined as:
\begin{equation}
m_{kl}=\left\langle \sigma_{k}\sigma_{l}\right\rangle =\frac{1}{Z_{N}\left(\left\{ K_{ij}\right\} \right)}\sum_{\left\{ \sigma_{i}=\pm1\right\} }\sigma_{k}\sigma_{l}\exp\left(\frac{1}{2}\sum_{i,j}^{N}K_{ij}\sigma_{i}\sigma_{j}\right)=-\frac{1}{T}\frac{\partial\mathcal{F}_{N}\left(\left\{ K_{ij}\right\} \right)}{\partial K_{kl}}\,.\label{eq:150}
\end{equation}
The notation $\left\langle \dots\right\rangle $ is used for \emph{thermodynamic
average} with Boltzmann statistical weights $\exp\left(-{\cal H}/T\right)$.
The pair correlator gives us an information about spin ordering in
a system.

Thermodynamic quantities, namely the entropy and the heat capacity
may be written as: 
\begin{equation}
\mathcal{S}_{N}\left(\left\{ K_{ij}\right\} \right)=-\frac{\partial\mathcal{F}_{N}\left(\left\{ K_{ij}\right\} \right)}{\partial T}=\frac{1}{2T}\sum_{k,l}K_{kl}\frac{\partial\mathcal{F}_{N}\left(\left\{ K_{ij}\right\} \right)}{\partial K_{kl}}=-\frac{1}{2}\sum_{k,l}K_{kl}m_{kl}\,,\label{eq:160}
\end{equation}
\begin{equation}
\mathcal{C}_{N}\left(\left\{ K_{ij}\right\} \right)=T\frac{\partial\mathcal{S}_{N}\left(\left\{ K_{ij}\right\} \right)}{\partial T}=-T\frac{\partial^{2}\mathcal{F}_{N}\left(\left\{ K_{ij}\right\} \right)}{\partial T^{2}}\,.\label{eq:170}
\end{equation}
respectively.

\subsubsection{Thermodynamic limit and phase transitions.\label{sub:limit}}

In the rest of the paper the limit of infinite system size, $N\to\infty$,
will be implied everywhere. Therefore, the specific values per spin
are to be introduced for extensive quantities $\sim N$: for free
energy, 
\begin{equation}
f\left(\left\{ K_{ij}\right\} \right)=\lim_{N\to\infty}\frac{\mathcal{F}_{N}\left(\left\{ K_{ij}\right\} \right)}{N}\,,\label{eq:180}
\end{equation}
for the entropy,
\begin{equation}
s\left(\left\{ K_{ij}\right\} \right)=\lim_{N\to\infty}\frac{\mathcal{S}_{N}\left(\left\{ K_{ij}\right\} \right)}{N}=-\frac{\partial f\left(\left\{ K_{ij}\right\} \right)}{\partial T}=\frac{1}{2T}\sum_{k,l}K_{kl}\frac{\partial f\left(\left\{ K_{ij}\right\} \right)}{\partial K_{kl}}\label{eq:190}
\end{equation}
and for the heat capacity,
\begin{equation}
c\left(\left\{ K_{ij}\right\} \right)=T\frac{\partial s\left(\left\{ K_{ij}\right\} \right)}{\partial T}=-T\frac{\partial^{2}f\left(\left\{ K_{ij}\right\} \right)}{\partial T^{2}}\label{eq:200}
\end{equation}

Let us consider the Ising model on an infinite $D$-dimensional hypercubic
lattice. The lattice constant is set to be equal to $1$. Assume all
$J_{ij}$ to be equal, $J_{ij}=J>0$. In a hypercubic lattice due
to its homogeneity the correlator is either $m_{ij}=m\left(r_{ij},T\right)$
if $J>0$, or $m_{ij}=\pm m\left(r_{ij},T\right)$ if $J<0$ ($+$if
spins $i$ and $j$ are in the same sublattice and $-$otherwise),
$m$ depends only on the distance between spins $r_{ij}=\left|\boldsymbol{r}_{i}-\boldsymbol{r}_{j}\right|$and
on temperature $T$. There exists some temperature point $T_{c}$,
called critical temperature. The correlator at large distances decays
exponentially either to zero or to some $m^{2}\left(T\right)$, $0<m^{2}\left(T\right)<1$:
\begin{equation}
m\left(r,T\right)\to m^{2}\left(T\right)\left[\theta\left(T_{c}-T\right)+\exp\left(-\frac{r}{l_{c}\left(T\right)}\right)\right]\,\textrm{as\;r\ensuremath{\to\infty}},\label{eq:210}
\end{equation}
where $l_{c}\left(T\right)$ is the \emph{correlation length}, $l_{c}\left(T\right)\to\infty$
as $T\to T_{c}$ and $m\left(T\right)$ is the \emph{spontaneous magnetization},
--- the \emph{order parameter} for the ferromagnetic or antiferromagnetic
phase transition.

The thermodynamic quantities: free energy, entropy and specific heat
as a functions of temperature have singularities at $T=T_{c}$ as
well as an order parameter. At $T\to0$, the order parameter tends
to $1$, the entropy and specific heat tend to zero as $\exp\left(-2DJ/T\right)$.

\subsection{Configuration average.\label{sub:disordered}}

In this chapter I will assume, just for definiteness, the graph to
be the $D$-dimensional cubic lattice with periodic boundary conditions.
The interactions are between nearest neighbors only. The number of
sites is $N=L^{D}$. Coupling energies $J_{ij}$ assumed to be independent
and statistically equivalent random variables. Their statistical properties
are completely defined by e.g. the probability density
\begin{equation}
p\left(j\right)=\overline{\delta\left(J_{ik}-j\right)}\,,\label{eq:220}
\end{equation}
where $\overline{\left.\dots\right.}$ means a \emph{configuration
average}, as opposed to a thermodynamic average $\left\langle \dots\right\rangle $.

The free energy and its temperature derivatives, entropy and heat
capacity are a self-averaging quantities, i.e. in the thermodynamic
limit we have: $\overline{\left(f-\overline{f}\right)^{2}}=0$ etc.
The value of the pair correlator $m_{ij}$ is fluctuating. Its statistical
properties may be completely described, for example, by the set of
momenta momenta:
\begin{equation}
{\cal Q}_{p}\left(r_{kl}\right)=\overline{m_{kl}^{p}}=\overline{\left\langle \sigma_{k}\sigma_{l}\right\rangle ^{p}}\,,p=0,1,2,\dots.\label{eq:225}
\end{equation}

Free energy now is a \emph{functional of the distribution }of\emph{
$K_{ij}=J_{ij}/T$.} For example, one can write \emph{$f=\mathrm{f}\left(\left\{ P\left(k\right)\right\} \right)$},
where\emph{
\begin{equation}
P\left(k\right)=\overline{\delta\left(K_{ij}-k\right)}\,.\label{eq:230}
\end{equation}
}For our purposes two another representations of the real random variable
$K$ are more convenient, namely Fourier (or two-sided Laplace) representation:
\begin{equation}
\Phi\left(\xi\right)=L\left(i\xi\right)=\overline{e^{i\xi K}}=\int_{-\infty}^{+\infty}dk\,P\left(k\right)e^{i\xi k}\,,\;\xi\in\mathbb{C}\label{eq:240}
\end{equation}
and the ${\cal T}$-representation, defined as:
\begin{equation}
{\cal T}_{q}=\overline{\tanh^{q}K}=\int_{-\infty}^{+\infty}dk\,P\left(k\right)\tanh^{q}k\,,\;q=0,1,2,\dots\label{eq:250}
\end{equation}
They are related as follows:
\begin{equation}
\Phi\left(\xi\right)=L\left(i\xi\right)=1+i\xi\sum_{q=1}^{\infty}S_{q-1}\left(i\xi\right){\cal T}_{q}\,,\label{eq:260}
\end{equation}
\begin{equation}
{\cal T}_{q}=-\frac{i}{2q}\int_{-\infty+i0}^{+\infty+i0}\frac{d\xi}{\sinh\left(\pi\xi/2\right)}S_{q-1}\left(-i\xi\right)\Phi\left(\xi\right)=\frac{i}{2q}\int_{-i\infty-0}^{+i\infty-0}\frac{dx}{\sin\left(\pi x/2\right)}S_{q-1}\left(-x\right)L\left(x\right)\,,\label{eq:270}
\end{equation}
where $S_{q}\left(x\right)\,,q=0,1,2,\dots$ are polynomials, defined
by the following formula:
\begin{equation}
e^{xy}=1+x\sum_{q=0}^{\infty}S_{q}\left(x\right)\tanh^{q+1}y\,.\label{eq:280}
\end{equation}
They form a complete orthogonal set on an imaginary $x$-axis (see
Appendix \ref{sub:polynomials} for details).

Eq. (\ref{eq:260}) follows directly from Eqs. (\ref{eq:280}) and
(\ref{eq:250}), and Eq. (\ref{eq:270}) --- from Eq. (\ref{eq:280})
and orthogonality condition (\ref{eq:a140}).

The free energy $f$ may be considered as a functional of $\Phi\left(\xi\right)$,
$f=\mathrm{f}\left(\left\{ \Phi\left(\xi\right)\right\} \right)$.
Let us choose arbitrary a site, say $k$ and add one extra interaction
with dimensionless strength $\varkappa$ between spin $\sigma_{k}$
and some other one, $\sigma_{l}$. It means, that the new value the
partition function is:
\[
Z_{N}\left(\left\{ K_{ij}\right\} ;\varkappa,k,l\right)=\cosh\varkappa\left(1-m_{kl}\tanh\varkappa\right)Z_{N}\left(\left\{ K_{ij}\right\} \right)\,.
\]
The free energy in the thermodynamic limit changes as $f\to f+\delta f\left(r,\varkappa\right)$,
$\delta f\left(r,\varkappa\right)=-\left(T/N\right)\left[\ln\cosh\varkappa+\overline{\ln\left(1-m_{kl}\tanh\varkappa\right)}\right]$,
$r\equiv r_{kl}$ is the distance between sites $k$ and $l$. (\ref{eq:225})
\begin{equation}
\delta f\left(r,\varkappa\right)=-\frac{T}{N}\sum_{q=1}^{\infty}\left[\frac{1+\left(-1\right)^{q}}{2}-{\cal Q}_{q}\left(r\right)\right]\frac{\tanh^{q}\varkappa}{q}\label{eq:290}
\end{equation}

\section{Real space renormalization group\label{sec:RSRG} }

The main idea of a renormalization group is an elimination of some
degrees of freedom of a system under consideration. In the statistical
mechanics of equilibrium systems this means partial summation in a
partition function of a system. After such a procedure one arrives
at a \emph{renormalized} system with a new Hamiltonian. Equations,
connecting the parameters of the renormalized Hamiltonian with the
ones of the original Hamiltonian are named renormalization equations.
Usually, after writing them in the infinitesimal limit, one arrives
at a set of equations, describing evolution of a system's parameters
with a growing lengthscale.

For the Ising model the real space renormalization means the partial
summation in the Eq. (\ref{eq:110}). For example, let us consider
the Ising model on a 2D square lattice. The latter may be subdivided
in two sublattices with lattice constants equal to $\sqrt{2}$. After
summing in Eq. (\ref{eq:110}) of the spins of one sublattice one
arrives at the renormalized Hamiltonian. But it is not similar to
the original one, Eq. (\ref{eq:100}). Additionally, it contains terms,
which looks like $J_{ijkl}\sigma_{i}\sigma_{j}\sigma_{k}\sigma_{l}$.
To avoid this kind of difficulties, it was suggested to consider the
Ising model on a \emph{hierarchical graph}.

\subsection{RSRG on a hierarchical graph\label{sub:hierarchical}}

\begin{figure}
\hfill{}\subfloat[\label{fig:Formation}Formation of hierarchical graph.]{

\includegraphics[width=0.7\textwidth]{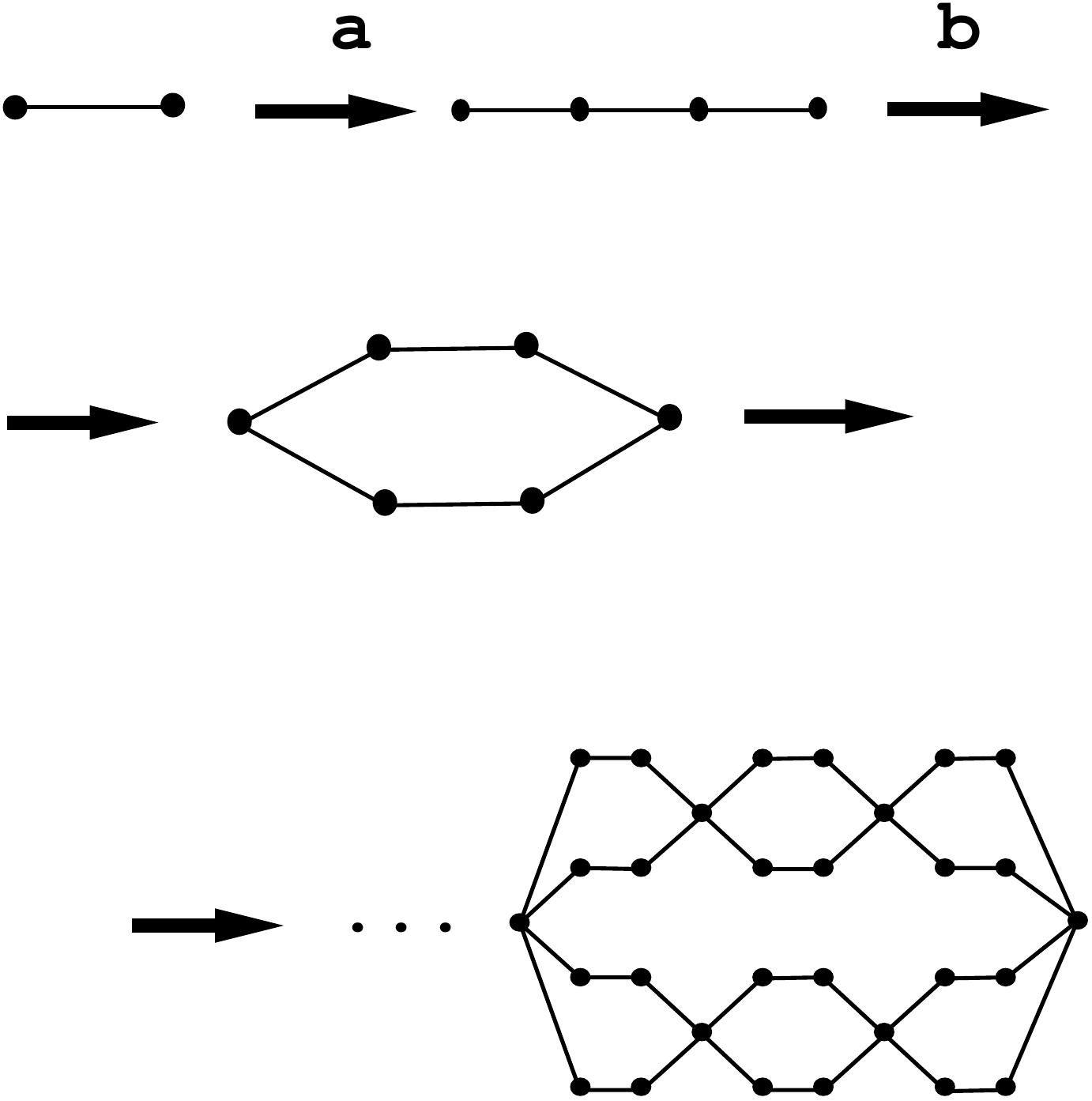}}\hfill{}\subfloat[\label{fig:graph2d}The same graph embedded in 2D. Bold vertical lines
correspond to infinite and positive interaction energies. ]{

\includegraphics[width=0.7\textwidth]{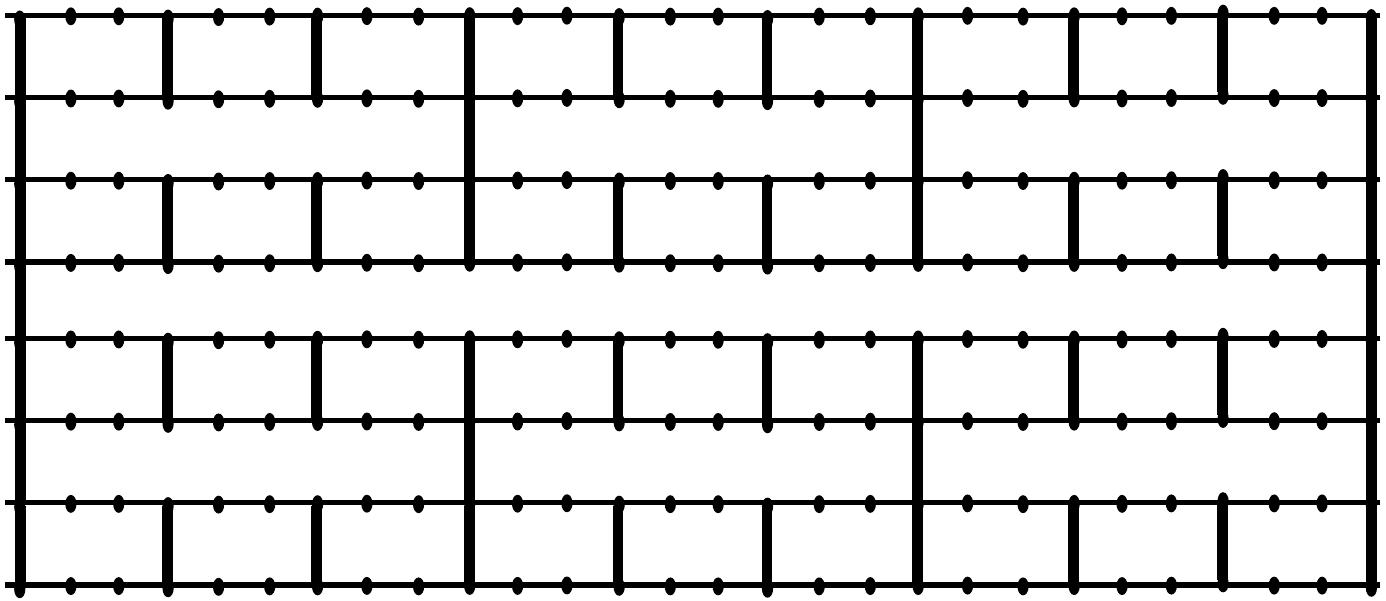}}\hfill{}

\caption{\label{fig:graph}Example of hierarchical graph of fractional dimensionality
$D=1+\ln2/\ln3$.}
\end{figure}

The latter is constructed as follows: Let us take two vertices and
a link, connecting them. Take $n$ copies of this link and connect
them, forming the chain of length $n$. Then take $m$ copies of the
chain and connect them in parallel. Thus, the $\left(n,m\right)$
hierarchical graph of the level $1$ is formed. Now repeat the procedure,
taking the resulting $\left(n,m\right)$ bundle (graph of the level
$1$) as an initial element (see Fig. \ref{fig:graph}a). And so on.
The Ising spins are placed on the vertices of the graph.

The $\left(n,m\right)$ hierarchical graph has fractional dimensionality,
equal to $D=1+\ln m/\ln n$. For example, at the Fig. \ref{fig:graph}b
the same graph as on the Fig. \ref{fig:graph}a is presented in a
different way: in 2D it it looks like a set of 1D chains connected
by the infinite strength links (bold vertical lines). Inside a square
of the size $3^{l}$ the chains may be subdivided into the bundles
of the widths $\le2^{l}$, therefore the Hausdorff dimensionality
in this example is equal to $D=1+\ln2/\ln3\approx1.631\dots$. Generally,
an $\left(n,m\right)$ hierarchical graph may be represented as a
set of 1D chains embedded in the $D_{e}$-dimensional space, $D_{e}$
is an integer and $\ln m/\ln n<D_{e}<1+\ln m/\ln n$. The chains in
a bundle are all interconnected by infinite strength bonds, and the
area of a bundle is $\mathcal{M}\le\left(\mathrm{size}\right)^{D-1}$,
therefore: 
\begin{equation}
m=n^{D-1}\,,D=1+\frac{\ln m}{\ln n}\,.\label{eq:310}
\end{equation}

The renormalization procedure on $\left(n,m\right)$ hierarchical
graph is described in Appendix \ref{sec:transformation}. It results
in:
\begin{enumerate}
\item The transformation of the distribution of the dimensionless interaction
energies $K$, $\left\{ {\cal T}_{q}^{\left(l\right)},\Phi_{l}\left(\xi\right)\right\} \to\left\{ {\cal T}_{q}^{\left(l+1\right)},\Phi_{l+1}\left(\xi\right)\right\} $,where
$l$ is the number of the renormalization in their sequence. It is:
\begin{equation}
\Phi_{l+1}\left(\xi\right)=\tilde{\Phi}_{l}^{m}\left(\xi\right);\;\tilde{{\cal T}}_{q}^{\left(l\right)}=\left[{\cal T}_{q}^{\left(l\right)}\right]^{n}\,,\label{eq:320}
\end{equation}
where $\tilde{{\cal T}}_{q}^{\left(l\right)}$ and $\tilde{\Phi}_{l}\left(\xi\right)$
are ${\cal T}$- and Fourier representations of the distribution of
the effective couplings $\tilde{K}^{\left(l\right)}$ for 1D chains.
\item Formula for the free energy:
\begin{equation}
\mathrm{f}=T\sum_{l=0}^{\infty}\left(mn\right)^{-l}\left[\frac{1}{n}\overline{\ln\left(2\cosh\tilde{K}^{\left(l\right)}\right)}-\overline{\ln\left(2\cosh K^{\left(l\right)}\right)}\right]\label{eq:330}
\end{equation}

\end{enumerate}

\subsection{Infinitesimal limit\label{sub:Infinitezimal}}

Real space $\left(n,m\right),\,m=n^{D-1}$ renormalization on a hierarchical
graph mimics $n$ times length rescaling on a $D$-dimensional lattice.
Let us make an analytic continuation on $n$, and then set $n=1+d\lambda/\lambda,\,m=1+\left(D-1\right)d\lambda/\lambda$.
The parameter $\lambda=n^{l}$ is the length scale. ${\cal T}_{q}^{\left(l\right)}$
and $\Phi_{l}\left(\xi\right)$ are to be replaced with ${\cal T}_{q}\left(\lambda\right)$
and $\Phi\left(\lambda,\xi\right)$, resp. Then from Eq. (\ref{eq:320}),
taking into account Eqs. (\ref{eq:260},\ref{eq:270}) I arrive at
the renormalization equations, which can be written as:
\begin{gather}
\lambda\frac{d{\cal T}_{q}\left(\lambda\right)}{d\lambda}={\cal T}_{q}\left(\lambda\right)\ln{\cal T}_{q}\left(\lambda\right)-\frac{i\left(D-1\right)}{2q}\int_{-\infty}^{+\infty}\frac{d\xi}{\sinh\left(\pi\xi/2\right)}S_{q-1}\left(-i\xi\right)\Phi\left(\lambda,\xi\right)\ln\Phi\left(\lambda,\xi\right)\,,\nonumber \\
\Phi\left(\lambda,\xi\right)=1+i\xi\sum_{q=1}^{\infty}S_{q-1}\left(i\xi\right){\cal T}_{q}\left(\lambda\right)\,,\label{eq:340}
\end{gather}
or, alternatively, as:
\begin{gather}
\lambda\frac{\partial\Phi\left(\lambda,\xi\right)}{\partial\lambda}=\left(D-1\right)\Phi\left(\lambda,\xi\right)\ln\Phi\left(\lambda,\xi\right)+i\xi\sum_{q=1}^{\infty}S_{q-1}\left(i\xi\right){\cal T}_{q}\left(\lambda\right)\ln{\cal T}_{q}\left(\lambda\right)\,,\nonumber \\
{\cal T}_{q}\left(\lambda\right)=-\frac{i}{2q}\int_{-\infty+i0}^{+\infty+i0}\frac{d\xi}{\sinh\left(\pi\xi/2\right)}S_{q-1}\left(-i\xi\right)\Phi\left(\xi\right)\,.\label{eq:350}
\end{gather}

\subsection{Free energy and correlator momenta\label{sub:Free}}

\subsubsection{Free energy,entropy and heat capacity\label{sub:thermodynamics}}

Free energy ${\rm f}$ is a functional of the dimensional couping
strengths $K_{ij}$ distribution. It may be written as $f\left(\left\{ {\cal T}_{q}\right\} \right)$or
as $f\left(\left\{ \Phi\left(\xi\right)\right\} \right)$, ${\cal T}_{q}\left(1\right)={\cal T}_{q}$
and $\Phi\left(1,\xi\right)=\Phi\left(\xi\right)$ are the initial
conditions for the Eqs. (\ref{eq:340}) and (\ref{eq:350}), resp.
From Eq. (\ref{eq:330}), using the following formula:
\begin{gather}
\overline{\ln\cosh K}=\sum_{q=1}^{\infty}\frac{{\cal T}_{2q}}{2q}\label{eq:360}
\end{gather}
I obtain:
\begin{equation}
f\left(\left\{ {\cal T}_{q}\right\} \right)=-T\frac{\ln2}{D}+T\int_{1}^{\infty}\frac{d\lambda}{\lambda^{D+1}}\sum_{q=1}^{\infty}\frac{\mathcal{T}_{2q}\left(\lambda\right)\left[\ln\mathcal{T}_{2q}\left(\lambda\right)-1\right]}{2q}\,.\label{eq:370}
\end{equation}

From the formula, alternative to (\ref{eq:360}),
\begin{equation}
\overline{\ln\left(2\cosh K\right)}=\frac{1}{2}\fintop_{-\infty}^{+\infty}d\xi\frac{1-\Phi\left(\xi\right)}{\xi\sinh\left(\pi\xi/2\right)}\,,\label{eq:380}
\end{equation}
where $\fintop$ is the principal value of the integral. This formula
may be easily derived using the inverse of Fourier Transform:
\[
P\left(k\right)=\int_{-\infty}^{+\infty}\frac{d\xi}{2\pi}\Phi\left(\xi\right)e^{-ik\xi}\,.
\]
 Then I get:
\begin{equation}
f\left(\left\{ \Phi\left(\xi\right)\right\} \right)=-\frac{\left(D-1\right)T}{2}\int_{1}^{\infty}d\lambda\,\lambda^{-D-1}\fint_{-\infty}^{+\infty}d\xi\frac{1+\Phi\left(\lambda,\xi\right)\left[\ln\Phi\left(\lambda,\xi\right)-1\right]}{\xi\sinh\left(\pi\xi/2\right)}\,,\label{eq:390}
\end{equation}

\subsubsection{Correlator momenta\label{sub:Correlator}}

Adding an extra link of dimensionless strength $\varkappa$ and of
length $r$ results in the (infinitesimal in the thermodynamic limit)
variation of $\Phi\left(r,\xi\right)$, $\Phi\left(r,\xi\right)\to\Phi\left(r,\xi\right)+\delta\Phi\left(r,\xi;\varkappa\right)$:
\begin{gather}
\delta\Phi\left(r,\xi;\varkappa\right)=\frac{r^{D}}{N}\left(e^{i\varkappa\xi}-1\right)\Phi\left(r,\xi\right)=\frac{r^{D}}{N}\sum_{q=1}^{\infty}\phi_{q}\left(r,\xi\right)\tanh^{q}\varkappa\,,\label{eq:400}\\
\phi_{q}\left(r,\xi\right)=\Phi\left(r,\xi\right)i\xi S_{q-1}\left(i\xi\right)\,,\label{eq:410}
\end{gather}
where Eq. (\ref{eq:a80}) was used. As $\lambda>r$, $\phi_{q}\left(\lambda,\xi\right)$
is to found from the following equations: 
\begin{gather}
\lambda\frac{\partial\phi_{q}\left(\lambda,\xi\right)}{\partial\lambda}=\left[D+\left(D-1\right)\ln\Phi\left(\lambda,\xi\right)\right]\phi_{q}\left(\lambda,\xi\right)++i\xi\sum_{q=1}^{\infty}\frac{S_{q-1}\left(i\xi\right)}{q}\ln\mathcal{T}_{q}\left(\lambda;\varkappa\right)t_{q}\left(\lambda\right)\,,\nonumber \\
t_{q}\left(\lambda\right)=-\frac{i}{2q}\int_{-\infty}^{+\infty}\frac{d\xi}{\sinh\left(\pi\xi/2\right)}S_{q-1}\left(-i\xi\right)\phi_{q}\left(\lambda,\xi;\varkappa\right)\,,\label{eq:420}
\end{gather}
with the initial condition at $\lambda=r$, given by Eq.(\ref{eq:410}).
They are obtained by the linearization of Eq. (\ref{eq:350}).

From Eq. (\ref{eq:390}) one can derive the following relation for
the variation of the free energy:
\begin{gather}
\delta f\left(r,\varkappa\right)=-\frac{\left(D-1\right)T}{2}\int_{r}^{\infty}\frac{d\lambda}{\lambda^{D+1}}\int_{-\infty}^{+\infty}d\xi\frac{\ln\Phi\left(\lambda,\xi\right)\delta\Phi\left(\lambda,\xi;\varkappa\right)}{\xi\sinh\left(\pi\xi/2\right)}=\nonumber \\
=-\frac{\left(D-1\right)T}{2}\frac{r^{D}}{N}\sum_{q=1}^{\infty}\tanh^{q}\varkappa\int_{r}^{\infty}\frac{d\lambda}{\lambda^{D+1}}\int_{-\infty}^{+\infty}d\xi\frac{\ln\Phi\left(\lambda,\xi\right)\phi_{q}\left(\lambda,\xi\right)}{\xi\sinh\left(\pi\xi/2\right)}.\label{eq:430}
\end{gather}
Comparing Eq. (\ref{eq:430}) with Eq. (\ref{eq:290}), I get the
expression for the correlator momenta:
\begin{equation}
{\cal Q}_{q}\left(r\right)=\frac{1+\left(-1\right)^{q}}{2}-\frac{D-1}{2}qr^{D}\int_{r}^{\infty}\frac{d\lambda}{\lambda^{D+1}}\int_{-\infty}^{+\infty}d\xi\frac{\ln\Phi\left(\lambda,\xi\right)\phi_{q}\left(\lambda,\xi\right)}{\xi\sinh\left(\pi\xi/2\right)}\,.\label{eq:440}
\end{equation}

\section{Conclusions\label{sec:Conclusions}}

As a result, we have an expressions for the free energy and its temperature
derivatives of a disordered \emph{finite-dimensional} Ising model,
and the ones for a momenta of pair correlator as a function of interspin
distance and temperature. However, to obtain explicit expression,
one must:
\begin{enumerate}
\item To get the distribution of the effective interactions as a function
of the lengthscale $\lambda$. For this purpose it is necessary to
solve \emph{functional} renormalization equation, Eq. (\ref{eq:340}),
or, alternatively, Eq. (\ref{eq:350}). Then, using Eq. (\ref{eq:370})
(or Eq. (\ref{eq:390})), we obtain the free energy as a function
of temperature and other parameters.
\item To solve linearized equations, Eq. (\ref{eq:420}). Then, using Eq.
(\ref{eq:440}), it is possible to determine correlator momenta.
\end{enumerate}
The first task is the most complicated one. The equations are nonlinear,
the structures of the sets of their solutions may be very intricate
and there is very little hope to find the relevant solution in explicit
form. It is difficult to solve them numerically because of very slow
convergence. Therefore, it is necessary to look for a situations,
whenever these equations may be either simplified or replaced by a
solvable (or, least, treatable) approximations. After the first step
done, the second one is essentially easier because Eq. (\ref{eq:440})
is linear.

What information about spin structure can be extracted from the set
of correlator momenta? First, spontaneous magnetization (FM order
parameter) and magnetic susceptibility. Also, for the Edwards-Anderson
order parameter, which is $q_{EA}=\overline{\left\langle \sigma_{i}\right\rangle ^{2}}$,
we have:
\begin{equation}
q_{EA}^{2}=\lim_{r\to\infty}{\cal Q}_{2}\left(r\right)\,.\label{eq:450}
\end{equation}

Probably, also the Parisi functional order parameter can be somewhat
extracted from the set of momenta ${\cal Q}_{2q}\left(r\right)$.
This question is still an open one for the author.

\section{Announcement of the next paper in series\label{sec:Announcement}}

In the second paper in the series the RSRG equations will be considered
in the low-temperature limit, $T\to0$. There we shall concentrate
on the symmetric distributions, $P\left(k\right)=P\left(-k\right)$.
In this case only even momenta of correlator are nonzero, ${\cal Q}_{2q+1}=0$.
Then the rescaling equations simplify drastically and can be investigated
analytically and numerically. As a result, it is possible to prove,
that, independent of the specific choise of the exchange energies
distribution, we arrive at the universal form of the effective couplings
distribution, $P\left(\lambda,k;T\right)\to\left(\lambda^{\alpha}/T\right)\overline{P}\left(Tk/\lambda^{\alpha}\right)$
as $\lambda\apprge1$. Here both the functional form of $\overline{P}$
and the value of the scaling index $\alpha$ depend only on the system's
dimensionality $D$. At the moment some statements may be claimed:
\begin{enumerate}
\item The lower critical dimensionality is equal to $D_{l}=2.4929\dots\ne5/2$.
The scaling index $\alpha\left(D\right)$, being analytic function
of $D$, $D>1$ changes sign at $D=D_{l}$, from negative at $D<D_{l}$
to positive at $D>D_{l}$. The spin glass phase is possible only if
$D>D_{l}$. The critical temperature $T_{g}\to0$ as $D\to D_{l}+0$.
\item The residual entropy is zero: $s\left(T\right)\to0$ as $T\to0$.
\item The heat capacity in the low-temperature limit is also zero. However,
$c\left(T\right)$ has a weak singularity at $T=0$:
\begin{equation}
c\left(T\right)\sim aT+bT^{2}\ln\frac{1}{T}\,.\label{eq:520}
\end{equation}

\item Correlation's momenta may be written in a scaling form, ${\cal Q}_{2q}\left(r,T\right)=\overline{{\cal Q}}_{2q}\left(r^{\alpha}/T\right)$,
but this still questionable to me moment As $D-D_{l}\ll1$, there
is a kind of hyperscaling: all the relevant characteristics of the
system are contained in only one function, say in the stable critical
probability density for the effective interaction energies $\overline{P}$.
It is calculated numerically, and its asymptotic behaviors in various
limits are analyzed.\end{enumerate}
\begin{acknowledgments}
I am very grateful to the participants of the theoretical seminar
at the department of Physics of the University of Aveiro, Portugal,
where this work was started, and to the participants of the seminar
of the Sector of Theory of Semiconductors and Dielectrics of the Ioffe
Institute, where the work was finished. The author also would like
to thank Prof. J-F.F. Mendes for his hospitality and patience during
the author's stay at the University of Aveiro. Special thanks to Prof.
Sergey Dorogovtsev..
\end{acknowledgments}

\appendix

\section{Polynomials $S_{q}\left(x\right)$.\label{sub:polynomials}}

They may be defined, for example, using a generating function:
\begin{equation}
G\left(x,z\right)\equiv\frac{1}{xz}\left[\left(\frac{1+z}{1-z}\right)^{x/2}-1\right]=\sum_{q=0}^{\infty}S_{q}\left(x\right)z^{q}\,.\label{eq:a70}
\end{equation}
Setting in Eq. (\ref{eq:a70}) $z=\tanh y$, we get:
\begin{equation}
e^{xy}=1+x\sum_{q=0}^{\infty}S_{q}\left(x\right)\tanh^{q+1}y\,.\label{eq:a80}
\end{equation}

$S_{q}\left(x\right)$ may be represented as a contour integral
\begin{equation}
S_{q}\left(x\right)=\frac{1}{x}\ointctrclockwise_{\left|z\right|<1}\frac{dz}{2\pi i}z^{-q-2}\left(\frac{1+z}{1-z}\right)^{x/2}\,.\label{eq:a90}
\end{equation}
Changing in Eq. (\ref{eq:a90}) the integration variable, $z=\tanh y$
and integrating by parts, we have:
\begin{equation}
S_{q}\left(x\right)=\frac{1}{q+1}\ointctrclockwise_{C}\frac{dy}{2\pi i}e^{xy}\coth^{q+1}y\,,\label{eq:a95}
\end{equation}
where the integration contour $C$ encircles the singularity point
$y=0$ anticlockwise.

The following integral representation:
\begin{equation}
S_{q}\left(x\right)=\frac{\sin\left(\pi x/2\right)}{\pi x}\int_{-\infty}^{+\infty}dk\,e^{xk}\frac{\tanh^{q}k}{\cosh^{2}k}\label{eq:a97}
\end{equation}
may be obtained if to integrate by parts in Eq. (\ref{eq:a95}) and
choose the integration contour as:
\[
C=\left(-\infty-i\pi/2,+\infty-i\pi/2,+\infty+i\pi/2,-\infty+i\pi/2,-\infty-i\pi/2\right)\,.
\]

The recurrence relation, 
\begin{equation}
xS_{q}\left(x\right)=\left(q+2\right)S_{q+1}\left(x\right)-qS_{q-1}\left(x\right)\,,\label{eq:a100}
\end{equation}
immediately follows from Eq. (\ref{eq:a95}) after integration by
parts:
\[
xS_{q}\left(x\right)=\frac{x}{q+1}\ointctrclockwise_{C}\frac{dy}{2\pi i}e^{xy}\coth^{q+1}y=\ointctrclockwise_{C}\frac{dy}{2\pi i}e^{xy}\frac{\coth^{q}y}{\sinh^{2}y}=\ointctrclockwise_{C}\frac{dy}{2\pi i}e^{xy}\coth^{q}y\left(\coth^{2}y-1\right)
\]

$S_{q}\left(x\right)$ may be expressed through the Gaussian hypergeometric
function $_{2}F_{1}$as follows: Changing the integration variable
in Eq. (\ref{eq:a90}) as $z\to\zeta=1/z$, one arrives (after the
proper deformation of the integration contour) at the following formula:
\[
S_{q}\left(x\right)=\frac{\sin\left(\pi x/2\right)}{\pi x}\int_{-1}^{1}d\zeta\,\zeta^{q}\left(\frac{1+\zeta}{1-\zeta}\right)^{x/2}\,.
\]
Introducing the symmetric function of two complex variables,
\begin{gather}
\text{\ensuremath{\mathit{\Sigma}}}\left(q,x\right)=q\int_{0}^{1}dz\,z^{q-1}\left(\frac{1-z}{1+z}\right)^{x}=2q\int_{0}^{\infty}dk\,e^{-2qk}\tanh^{x}k=\nonumber \\
=\frac{\Gamma\left(1+q\right)\Gamma\left(1+x\right)}{\Gamma\left(1+q+x\right)}\,_{2}F_{1}\left(x,q;1+q+x;-1\right)=\text{\ensuremath{\mathit{\Sigma}}}\left(x,q\right)\,,\label{eq:a110}
\end{gather}
we get:
\begin{equation}
S_{q}\left(x\right)=\frac{\sin\left(\pi x/2\right)}{\pi x}\left[\mathit{\Sigma}\left(q+1,-x/2\right)+\left(-1\right)^{q}\mathit{\Sigma}\left(q+1,x/2\right)\right]\,.\label{eq:a120}
\end{equation}

Polynomials $S_{q}\left(i\xi\right)=S_{q}^{*}\left(-i\xi\right),\,q=0,1,2,\dots$
form a complete orthogonal set of functions on $\left(-\infty,+\infty\right)$
with the weight function: 
\begin{equation}
W\left(\xi\right)=\frac{\xi}{2\sinh\left(\pi\xi/2\right)}\,.\label{eq:a130}
\end{equation}

\begin{itemize}
\item \textbf{Orthogonality}.
\begin{equation}
\int_{-\infty}^{+\infty}d\xi\,W\left(\xi\right)S_{p}^{*}\left(i\xi\right)S_{q}\left(i\xi\right)=\frac{\delta_{pq}}{p+1}\,.\label{eq:a140}
\end{equation}
This equation may be easily derived using the integral representation
(\ref{eq:a95}) and the following formula:
\[
\int_{-\infty}^{+\infty}d\xi W\left(\xi\right)e^{i\left(y_{1}-y_{2}\right)\xi}=\frac{1}{\cosh^{2}\left(y_{1}-y_{1}\right)}=\frac{1}{\cosh^{2}y_{1}\cosh^{2}y_{2}}\sum_{r=0}^{\infty}\left(r+1\right)\coth^{-r}y_{1}\coth^{-r}y_{2}\,.
\]

\item \textbf{Completeness.
\begin{equation}
\sum_{q=0}^{\infty}\left(q+1\right)S_{q}^{*}\left(i\xi_{1}\right)S_{q}\left(i\xi_{2}\right)=\frac{\delta\left(\xi_{1}-\xi_{2}\right)}{W\left(\xi_{1}\right)}\,.\label{eq:a150}
\end{equation}
}This follows directly from the integral representation (\ref{eq:a97})
using
\[
\sum_{q=0}^{\infty}\left(q+1\right)\tanh^{q}k_{1}\tanh^{q}k_{2}=\frac{\cosh^{2}k_{1}\cosh^{2}k_{2}}{\cosh^{2}\left(k_{1}-k_{2}\right)}\,.
\]

\end{itemize}

\section{Renormalization transformation on a regular hierarchical graph\label{sec:transformation}}

The partition function of the Ising model may be written as:
\begin{equation}
Z\left(\left\{ K_{ij}\right\} \right)=\sum_{\left\{ \sigma_{i}=\pm1\right\} }\prod_{i>j}T_{\sigma_{i},\sigma_{j}}\left(K_{ij}\right)\,.\label{eq:b10}
\end{equation}
Here$\hat{T}\left(K\right)$ is $2\times2$ \emph{transfer matrix}
of a single bond with dimensionless interaction energy $K$. Its matrix
elements are defined as: $T_{\sigma,\sigma'}\left(K\right)=\exp\left(K\sigma\sigma'\right)$.
One can write:
\begin{equation}
\hat{T}\left(K\right)=\begin{pmatrix}e^{K} & e^{-K}\\
e^{-K} & e^{K}
\end{pmatrix}\,.\label{eq:b20}
\end{equation}

\begin{figure}
\includegraphics[bb=0bp 0bp 595bp 750bp,width=19cm]{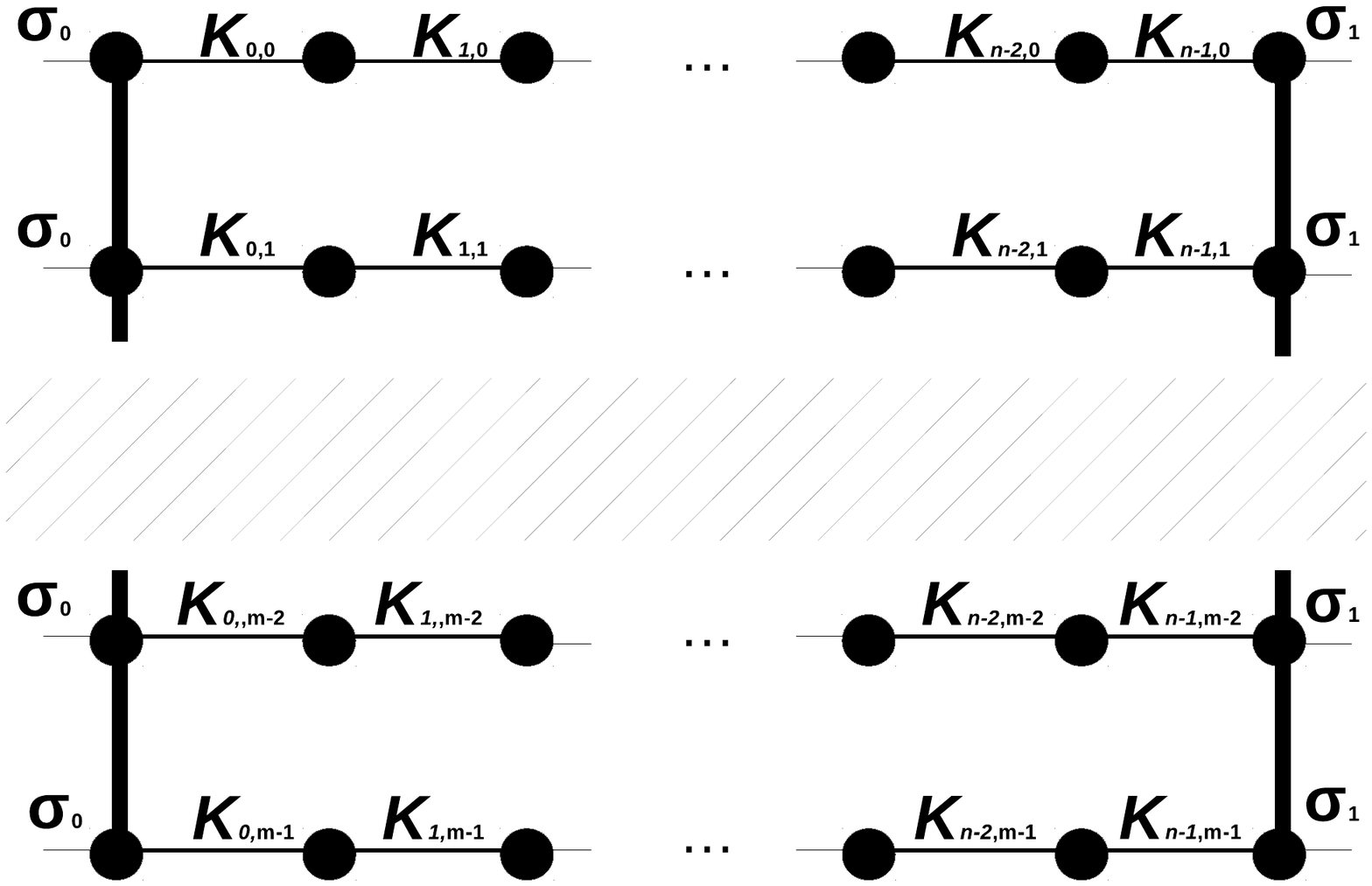}

\vspace{-11cm}
\caption{\label{fig:1-bond}Construction of the first level bond from$mn$
zero level ones. here the utmost left $m$ spins, connected by vertical
bold lines are $m$ copies of the spin$\sigma_{0},$the utmost right
ones --- of the spin $\sigma_{1}$.}
\end{figure}

Let us consider the part of the hierarchical graph, depicted on the
Fig. \ref{fig:1-bond}.

Let $\sigma_{0}$ be the top spin, and $\sigma_{1}$ --- the bottom
one on the figure.e main procedure in the RSRG method is the one of
partial summation in the partition function. Here it means the summation
over all the spins, except $\sigma_{0}$ and $\sigma_{1}$. As a result,
we get the contribution to the renormalized statistical weight of
the model, which can be written as:
\begin{gather}
Z\left(\sigma_{0},\sigma_{1}\right)=\prod_{j=0}^{m-1}\left[\prod_{i=0}^{n-1}\hat{T}\left(K_{i,j}\right)\right]_{\sigma_{0}\sigma_{1}}=\prod_{j=0}^{m-1}\tilde{T}_{\sigma_{0}\sigma_{1}}^{\left(j\right)}\,,\label{eq:b30}
\end{gather}
One can easily ascertain, that the transfer matrix $\hat{\tilde{T}}$$^{\left(j\right)}$
of the $j$-th chain may be expressed as: 
\begin{gather}
\hat{\tilde{T}}^{\left(j\right)}=\prod_{i=0}^{n-1}\hat{T}\left(K_{i,j}\right)=\frac{1}{2\cosh\tilde{K}_{j}}\prod_{i=0}^{n-1}\left(2\cosh K_{i,j}\right)\hat{T}\left(\tilde{K}_{j}\right)\,,\label{eq:b40}
\end{gather}
where $\hat{T}$ is defined in Eq. (\ref{eq:b20}), and the effective
interaction of the $j$-th chain $\tilde{K}_{j}$ is: 
\begin{equation}
\tanh\tilde{K}_{j}=\prod_{i=0}^{n-1}\tanh K_{ij}\,.\label{eq:b50}
\end{equation}
Then Eq. (\ref{eq:b30}) may be rewritten as:
\begin{equation}
Z\left(\sigma_{0},\sigma_{1}\right)=z\left(\left\{ K_{i,j}\right\} \right)\hat{T}_{\sigma_{0}\sigma_{1}}\left(K^{\left(1\right)}\right),\;\label{eq:b60}
\end{equation}
where
\begin{equation}
K^{\left(1\right)}=\sum_{j=0}^{m-1}\tilde{K}_{j}\,,\label{eq:b70}
\end{equation}
and
\begin{equation}
z\left(\left\{ K_{i,j}\right\} \right)=\left[\prod_{j=0}^{m-1}\left(2\cosh\tilde{K}_{j}\right)\right]^{-1}\left[\prod_{j=0}^{m-1}\prod_{i=0}^{n-1}\left(2\cosh K_{i,j}\right)\right]\label{eq:b80}
\end{equation}

As a result, we have the RSRG equations for dimensionless interactions
distribution,
\begin{equation}
\tilde{{\cal T}}_{q}=\overline{\prod_{i=0}^{n-1}\tanh^{q}K_{ij}}=\left[{\cal T}_{q}\right]^{n}\,,\label{eq:b90}
\end{equation}
\begin{equation}
\Phi_{1}\left(\xi\right)=\overline{\prod_{j=0}^{m-1}e^{i\xi\tilde{K}_{j}}}=\left[\tilde{\Phi}\left(\xi\right)\right]^{m}\,,\label{eq:b100}
\end{equation}
\begin{equation}
L_{1}\left(x\right)=\overline{\prod_{j=0}^{m-1}e^{x\tilde{K}_{j}}}=\left[\tilde{L}\left(x\right)\right]^{m}\,,\label{eq:b110}
\end{equation}
and the transformation formula for the free energy:
\begin{equation}
\bar{f}=T\left[\frac{1}{n}\overline{\ln\left(2\cosh\tilde{K}\right)}-\overline{\ln\left(2\cosh K\right)}\right]+\frac{1}{mn}\bar{f}_{1}\,.\label{eq:b120}
\end{equation}

\end{document}